\documentclass[useAMS,usenatbib]{mn2e}

\usepackage{graphicx}
\usepackage{times}
\usepackage{amsmath,amssymb}
\usepackage{epstopdf}

\newif\ifAMStwofonts
\AMStwofontstrue

\newcommand{\be}{\begin{equation}}
\newcommand{\ee}{\end{equation}}
\newcommand{\ba}{\begin{eqnarray}}
\newcommand{\ea}{\end{eqnarray}}
\newcommand{\brr}{\begin{array}}
\newcommand{\err}{\end{array}}
\newcommand{\bc}{\begin{center}}
\newcommand{\ec}{\end{center}}

\newcommand{\mincir}{\raise
 -2.truept\hbox{\rlap{\hbox{$\sim$}}\raise5.truept \hbox{$<$}\ }}
\newcommand{\magcir}{\raise
 -2.truept\hbox{\rlap{\hbox{$\sim$}}\raise5.truept \hbox{$>$}\ }}
\newcommand{\siml}{\raise
 -2.truept\hbox{\rlap{\hbox{$\sim$}}\raise5.truept \hbox{$<$}\ }}
\newcommand{\simg}{\raise
 -2.truept\hbox{\rlap{\hbox{$\sim$}}\raise5.truept \hbox{$>$}\ }}

\newcommand{\aap}{A\&A}
\newcommand{\apj}{ApJ}
\newcommand{\aj}{AJ}
\newcommand{\apjl}{ApJ}
\newcommand{\apjs}{ApJS}
\newcommand{\mnras}{MNRAS}

\newcommand{\prd}{Phys. Rev. D}

\title[]{Estimates of unresolved point sources contribution to WMAP 5} 
\author[L.P.L. Colombo et al.]{
L.P.L. Colombo$^1$ \& E. Pierpaoli$^1$,\\~\\ $^1$ University of Southern
California, Los Angeles, CA, 90089-0484\\}

\begin{document}

\date{Accepted ???. Received ???; in original form ???}

\maketitle

\begin{abstract} 

 We present an alternative estimate of the unresolved point source
 contribution to the WMAP temperature power spectrum based on current
 knowledge of sources from radio surveys in the $1.4-90$ GHz range.
 We implement a stochastic extrapolation of radio point sources in
 the NRAO-VLA Sky Survey (NVSS) catalog, from the original $1.4$ GHz
 to the $\sim 100$ GHz frequency range relevant for CMB
 experiments. With a bootstrap approach, we generate an ensemble of
 realizations that provides the probability distribution for the flux
 of each NVSS source at the final frequency. The predicted source
 counts agree with WMAP results for $S > 1$ Jy and the corresponding
 sky maps correlate with WMAP observed maps in Q-, V- and W- bands,
 for sources with flux $S > 0.2$ Jy. The low--frequency radio surveys
 found a steeper frequency dependence for sources just below the WMAP
 nominal threshold than the one estimated by the WMAP team. This
 feature is present in our simulations and translates into a shift of
 $0.3-0.4 \sigma$ in the estimated value of the tilt of the power
 spectrum of scalar perturbation, $n_s$, as well as $\omega_c$. This
 approach demonstrates the use of external point sources
 datasets for CMB data analysis.

\end{abstract}

\begin{keywords}
 cosmic microwave background - cosmology: observations - radio
 continuum: general - radio continuum: galaxies
\end{keywords}

\section{Introduction}\label{sec:intro}

Modern Cosmic Microwave Background (CMB) observations provide a
powerful test of our understanding of the Universe. Within the
generally accepted framework of a $\Lambda$CDM model, the 5-year
measurements by the Wilkinson Microwave Anisotropy Probe (WMAP)
constrain the fundamental cosmological parameters with a relative
accuracy of $1-10\%$ \citep{2009ApJS..180..306D,2009ApJS..180..330K},
while the upcoming observations by the Planck satellite are expected
to improve these numbers by at least a factor $\sim 3-4$
\citep[e.g.][]{2006astro.ph..4069T,2009MNRAS.398.1621C}. One of the
main scientific goals of Planck, and other proposed future CMB
missions \citep[e.g. EPIC,][]{2009arXiv0906.1188B}, is understanding
the nature of Inflation. Detection of the B-mode of CMB polarization
would provide direct evidence of a primordial background of
Gravitational Waves arising from Inflation
\citep{1997PhRvL..78.2058K,1997PhRvL..79.2180S}. Even without such
detection the CMB remains the most powerful probe of Inflation
currently accessible. A general prediction of inflationary models is
that the fractional amplitude of density fluctuations would be nearly
scale independent, so that the corresponding power spectrum could be
well approximated by an Harrison-Zeldovich form $P(k) \propto
k^{n_s}$, where the spectral index $n_s \simeq 1$. The amount of
deviation $n_s -1$ constrains the shape of the inflationary potential,
and current WMAP results already allow to rule out several models
\citep{2009ApJS..180..330K}. Significant improvements will come from
Planck and the next generation CMB missions.

However, exploitation of the full potential of CMB measurements
requires a deep understanding of instrumental systematics and a
careful cleaning of foreground contaminants. On small angular scales,
extragalactic point sources are an important source of
contamination. Bright sources, which can be detected with high
significance in CMB maps, are typically accounted for by masking a
small area around the source position during the estimation of the CMB
angular power spectra, $C_\ell$. On the other hand, to a first
approximation, undetected, and therefore unmasked, point sources
provide a Poisson noise contribution to the measured $C_\ell$ as well
as a non-Gaussian signature in the maps
\citep[e.g.][]{1998MNRAS.297..117T,2003ApJ...589...58P,2004MNRAS.354.1005P,2008PhRvD..77j7305S,2008PhRvD..77l3011B}. An
incorrect determination of this contamination can lead to relevant
biases on the estimated cosmological parameters, in particular on
$n_s$ \citep{2006ApJ...651L..81H,2008ApJ...688....1H}, whose accurate
measurement depends on a careful estimate of $C_\ell$ over the largest
range of scales probed by an experiment.  In this respect the greatest
expectations are now on Planck, which is a whole sky survey with a
small instrumental beam and low noise level.  If foregrounds and
systematics are well under control, Planck will be able to confirm or
falsify the WMAP findings about $n_s$ being significantly smaller than
one. As inflation predicts that the level of departure on $n_s$ from
one is proportional to the amount of gravitational waves expected,
this also has implications on the level of B-mode polarization signal
expected and is in turn relevant for the design and planning of future
missions. It is therefore appropriate to try to approach the issue of
point source contamination from different angles and conceive
different approaches to deal with it. In this paper, we undertake this
path with an application to WMAP data analysis which could lead to new
approaches for point sources subtraction for Planck.

At frequencies around $\sim 1$ GHz there is a wealth of information on
extragalactic sources, from nearly full-sky surveys with high angular
resolution and flux sensitivity, like the Green Bank Telescope GB6
survey \citep{1996ApJS..103..427G}, the NRAO-VLA Sky Survey
\citep[NVSS][]{1998AJ....115.1693C} or the Faint Images of the Radio
Sky \citep[FIRST][]{2003yCat.8071....0B}. On the other hand, at
frequencies $\sim 100$ GHz, corresponding to the optimal observational
window for CMB experiments, there are only few dedicated studies
covering a handful of sources. Most information comes from sources
detected in CMB experiments themselves, typically representing only
the bright end of the point source distribution. In addition, this
lack of information limited the possibility of extrapolating data from
low frequencies to the CMB {\em ``gold spot''}.

To estimate the residual point sources contribution, the WMAP team
assumed that unresolved sources followed the same frequency scaling as
the bright sources detected in the 5 year maps. The overall
normalization of the residual power spectrum was then determined by
a fit to maps at three highest frequencies. This approach can be
considered {\em internal}, as it takes into account (mostly) only the
WMAP data itself, without relying on the low-frequency information
form other available point sources surveys. As other radio surveys
provide alternative and further information on radio sources
properties, incorporating such results in the CMB data analysis
pipeline could improve the estimation of the extragalactic sources
contamination, or provide an {\em external} crosscheck of the WMAP
results.

Different authors provided estimates of source counts at CMB
frequencies on the basis of the physical properties of the different
source populations
\citep[e.g.][]{1998MNRAS.297..117T,2005A&A...431..893D}. However,
multifrequency studies, describing the scaling of sources from middle
frequencies, $\nu \sim 20$ GHz, to $\sim 100$ GHz, are now starting to
appear \citep{2008MNRAS.385.1656S} and more are expected in the
future. Together with existing studies describing sources' behaviour
from low to intermediate frequencies, these allow to predict sources
contamination in CMB maps starting from the low frequency surveys.
Such predictions could be used in future data analysis to make
specific assessments on point sources contamination in CMB maps in real
space, and account for them with techniques different from the
Fourier--based ones currently used.

In this work we follow the empirical approach outlined above to assess
the unresolved point sources contribution to the WMAP data and its
impact on cosmological parameters. We use the NVSS catalog as a
template of source fluxes and positions at radio frequency, and
extrapolate that information to CMB frequencies. At difference with
previous works, we extrapolate each individual source in the NVSS
rather than generating source populations at CMB frequencies by
randomly sampling a theoretical distribution, or by extrapolating only
the statistical properties of the source population
\citep[e.g.][]{2007MNRAS.379.1442W}.  This work therefore provides a
crosscheck of WMAP estimates of undetected point sources
contamination, and hence of the determination of $n_s$ and other
parameters. As mentioned above, since this approach maintains the
information on the actual spatial distribution of sources, it can be
useful in cleaning future Planck maps or to estimate the impact of
clustering.  Here, however, we refrain from using an alternative
likelihood approach to the treatment of point sources in the parameter
estimation and adopt the WMAP strategy to that aim.

The outline of the paper is as follow. In Section \ref{sec:ped} we
briefly review the contribution of unresolved sources to CMB
temperature spectra. Section \ref{sec:procedure} describes our
extrapolation procedure of NVSS source to WMAP frequencies. In section
\ref{sec:res} and \ref{sec:maps} we discuss the implication of our
approach at the number counts and maps level, respectively, while in
section \ref{sec:spectra} we debate the corresponding impact on
determination of cosmological parameters.  Finally, in section
\ref{sec:conc} we draw our conclusions.

\section{Unresolved Sources Contribution to CMB spectra} \label{sec:ped}

The angular resolutions of WMAP ($\gtrsim 13$ arcmin) and Planck
($\gtrsim 5$ arcmin) are significantly larger than the angular
dimensions of typical extragalactic radio sources, which can then be
effectively considered as point--like objects in the resulting CMB
maps. Several works studied the secondary contribution to CMB
anisotropies due to point sources below the detection threshold
\citep[e.g.][]{1989ApJ...344...35F,1996MNRAS.281.1297T,1998MNRAS.297..117T,1999A&A...346....1S}. In
this section we briefly summarize the main results.

Unresolved sources contribution to $C_\ell$ can be divided into a
Poisson term due to the random sources positioning, and a correction
accounting for clustering in the sources distribution. We consider CMB
observations at a single frequency $\nu_0$ and suppose that all
sources above a limit flux $S_c$ will be detected and masked from the
final map, and assume a source population characterized by the
differential counts $dN(>S)/dS$. 

The Poisson noise contribution to the measured CMB $C_\ell$ is then given by:
\begin{equation}
\label{eq:src}
C_\ell^{\rm src} =  g^2(\nu_0) \int_0^{S_c} dS \, S^2 \, \left| \frac{dN}{dS} \right|
\end{equation}
where $g(\nu)$ converts from flux density to thermodynamic temperature:
\begin{equation}
\label{eq:gnu}
g(\nu) = {\frac{c^2 h^2}{2k^3 T_0^2 x^2}} {\frac{(e^x -1)^2}{x^2 e^x}} ~, \, 
x = {\frac{h \nu}{k T_0}}~. 
\end{equation}
The source correction does not depend on $\ell$ and, since approximately
$C_\ell \propto \ell^{-2}$ at high multipoles up to $\ell \sim 2000$,
the source correction becomes relevant on angular scales smaller than $180/ \ell \sim 180/1000  \sim 0.2 \deg$. The
correction due to clustering is given by:
\begin{equation}
\label{eq:clust}
C_\ell^{clust} =  g^2(\nu_0) \omega_\ell \left( \int_0^{S_c} dS  \, S \, 
\left| \frac{dN}{dS} \right| \right)^2~;
\end{equation}
here $\omega_\ell $ is the harmonic transform of the source two-point 
angular correlation function.

\begin{figure}
\includegraphics[width=8.cm]{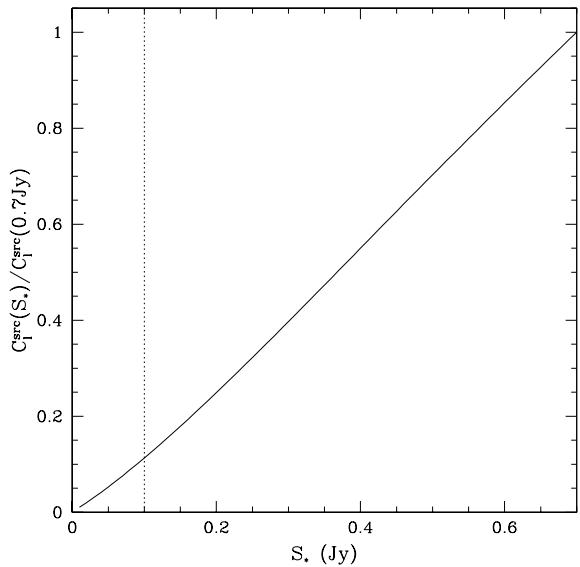}
\caption{Fraction of unresolved source contribution to WMAP spectra
 attributed to sources with fluxes $\le S_*$, assuming that all sources
 with $S \ge 0.7$ Jy have been removed from the maps. We used the
 de Zotti et al.(2005)  model for the number counts of sources
 with $S < 0.7$ Jy. Sources below $0.1$ Jy contribute $\lesssim 10\%$
 of the total unresolved power.}
\label{fig:lowflux}
\end{figure}

Real CMB experiments typically have complicated noise properties, due
to a combination of effects including instrumental systematics,
scanning strategy and foreground removal. As a consequence, the
probability of detecting a point source with flux $S_0$ will not be
uniform on the whole sky and $S_c$ will not be well defined. This is
not irrelevant as the unresolved point sources spectrum in
eq.\ref{eq:src} is typically dominated by the strongest amongst the
residual sources. While the source population above $ \gtrsim 1 $ Jy
in the WMAP 5 year catalog \citep[WMAP5,][]{2009ApJS..180..283W} is
well described by an Euclidean scaling of the number counts $dN/dS
\propto S^{-2.5}$, detected fainter sources suffers from significant
completeness issues. Depending on the frequency band considered,
$\gtrsim 90\%$ of detected sources have a flux $\ge 0.7$ Jy. Assuming
then an optimistic detection threshold $S_c = 0.7$ Jy and the
\cite{2005A&A...431..893D} model for $dN/dS$, we can estimate that
sources with $S \lesssim 0.1$ Jy contribute $\lesssim 10\%$ of the
total unresolved source power in WMAP data, as shown in
figure~\ref{fig:lowflux}. On the contrary, for Planck, whose $90 \%$
detection confidence limit is expected to be $S_c = 0.2$ Jy at 100 GHz
\citep{2003MNRAS.344...89V}, sources with $S \sim 0.1$ Jy will play a
dominant role in determination of $C_\ell^{src}$. Therefore, the
accuracy required for the extrapolations at a given flux level depends
on the characteristics of the target experiment.

\section{Extrapolation Procedure}\label{sec:procedure} 

\subsection{Reference Catalogs}\label{sec:catalogs}
Instead of extrapolating sources directly from NVSS to WMAP
frequencies, we consider a number of intermediate steps, depending on
the initial flux of the source considered. This is particularly
relevant at frequencies below 20 GHz, where a single power law is not
an accurate description for many sources
\citep[e.g.][]{2006MNRAS.371..898S}. By introducing several
intermediate steps, we account for the diversity of behaviours
observed in the data. The number and frequency range of the steps used
here was determined by available data. In selecting the reference
catalogs that we will use for the extrapolation, our goal was to get a
coverage of fluxes relevant for the estimation of unresolved sources
contribution to WMAP data over the full range of frequencies
considered. While this is not an issue in the $\sim $ GHz range,
coverage becomes sketchier and patchier at higher $\nu$. According to
the discussion in section \ref{sec:ped}, the main contamination to
WMAP spectra is due to undetected sources with fluxes $0.1 \sim 0.7$ Jy
in the CMB frequency range ($40-90$ GHz).
We will show this population to be dominated by sources with flat
spectrum ($\alpha > -0.5$ and $S \propto \nu^\alpha$), therefore
corresponding to sources with fluxes from a few hundreds of mJy to a
few  Jy at $1.4$ GHz.

In order to perform the extrapolation, we considered the following
sets of measurements:

\begin{itemize}

\item The NRAO-VLA Sky Survey (NVSS): it covers the full sky north of
 $-40\deg$ declination with an angular resolution of $45''$ at $1.4$ GHz. It
 includes about $1.8 \times 10^6$ discrete sources with a completeness limit
 of $S_{1.4} = 2.5$ mJy.

\item The Green Bank Telescope Survey at 6cm (GB6) observed the sky in
 the declination band $0\deg < \delta < 75\deg$ with an angular
 resolution $\sim 3.5'$ (however, the actual beam is not circular). It
 includes $75162$ sources above $S_{4.8} = 18$mJy.

\item The NVSS followup measurements by \cite{2009arXiv0901.4330M}.  A
 set of 3165 NVSS faint sources with $S_{1.4} < 100$ mJy and falling
 in the Cosmic Background Imager \citep[CBI][]{2004ApJ...609..498R}
 field was observed at $\nu = 31$ GHz using the Green Bank Telescope
 and the Owens Valley Radio Observatory. The resulting set of
 observations allowed to derive the distribution of spectral indexes
 $\alpha_{1.4}^{31}$ for sources with $S_{1.4} \le 0.1$ Jy.

\item The Ninth Cambridge Survey at 15 GHz (9C) followup observations
 by \cite{2007MNRAS.379.1442W}. The catalog includes 121 sources
 selected at $15.2$ GHz with a completeness limit of $S_{15} = 25$ mJy
 and simultaneous measurements at  $4.85$, $15.2$, $22$ and $43$ GHz.
 As the 9C survey covers the sky area observed by the Very Small Array 
\citep[VSA][]{2003MNRAS.341.1057W},
 which was selected to be devoid of bright sources, the sample contains only 
 a handful of sources with $S > 100$ mJy. 

\item The AT20G followup observations by \cite{2008MNRAS.385.1656S}.
 The work presents two samples of sources selected at $20$ GHz with
 simultaneous measurements at $90$ GHz: a first set of 59 inverted
 spectra sources with $S_{20} > 50$ mJy, and a flux-limited sample
 comprising 70 sources with $S_{20} >150$ mJy. We consider here only
 the flux-limited sample,  which constitutes our main reference sample
 for the extrapolation of sources with $S_{20} \le 1$ Jy from 20 to
 94  GHz.

\item At frequencies above $\sim 23 $  GHz and for fluxes above $\sim
 1$ Jy, the most comprehensive survey of point sources is provided by
 the WMAP 5 year catalog. WMAP observed the sky at five broad bands
 K, Ka, Q, V and W with central frequencies of $\sim 23, 33, 41 ,61,
 94$  GHz respectively. The catalog includes a total of 390 sources
 which have been detected with a $5\sigma$ confidence level in at
 least one of the bands, and is complete for fluxes above $\sim 1$ Jy
 in all bands.

\end{itemize}

\subsection{Method}\label{sec:methods}

Since the number of sources with high frequency information is smaller
than the number of NVSS sources, reconstructing the actual scaling of
individual sources is not possible. Our assumption is thus that from
the reference catalogs discussed above, we can build reference samples
which are representative of the behaviour of the whole source
population in the respective range of frequencies and fluxes. We then
perform a set of bootstrap simulations in which at each extrapolation
step, a simulated source is randomly paired to a reference source in
the catalog covering the next higher frequency range considered, and
we use the measured spectral index of the reference source to further
extrapolate the simulated source. Some of the catalogs considered
provide multifrequency information for their sources. When this is the
case, we use the set of spectral indexes of the catalog source to
extrapolate fluxes over the range of frequencies considered. For each
source in the NVSS catalog, we generate a set of 800
simulations. While in two different simulations a given source can be
extrapolated in different ways, the set of simulations provides the
probability distribution for flux at WMAP frequencies.

In broad terms, we define two main frequency ranges: 1) the
 interval from NVSS to the lowest WMAP band, $1.4-23$  GHz, and 2) the
 WMAP frequency range, $23-94$  GHz. Sources with $S_{1.4} \le 100$
 mJy are extrapolated directly to 23 GHz according to
 \cite{2009arXiv0901.4330M}, while for those with $S_{1.4} \ge 300$
 mJy we introduce a number of intermediate steps based based on the
 GB6 and \cite{2007MNRAS.379.1442W} measurements. In the intermediate
 flux range, we weight the two approaches in order to fit the 33  GHz
 low flux counts.  From 23  GHz upwards, sources with $S_{23} > 1$
 Jy are propagated based on the WMAP5 catalog, while for lower flux
 sources we base the extrapolation on the \cite{2008MNRAS.385.1656S}
 and \cite{2007MNRAS.379.1442W} reference samples. 

 In the following, we discuss each step in detail:

\begin{itemize}

\item {\it The 1.4 {\rm GHz} to 4.85 {\rm GHz} reference sample.}
  \cite{2008AJ....136..684K} provide a unified catalog of sources from
  four radio surveys NVSS, GB6, FIRST, WENS and from the optical SDSS
  surveys, matching sources in the different catalogs. From the
  NVSS-GB6 pairings identified in that work, we construct a reference
  sample of sources which provides spectral indexes between $1.4$ GHz
  and $4.85$ GHz. We consider only pairs in which the GB6 counterpart
  is found within $70''$ of the NVSS source, corresponding to an
  estimated completeness of $0.979$ and an efficiency (fraction of
  detected pairs corresponding to actual physical matches) of $0.79$
  \citep[for further details see][]{2008AJ....136..684K}. In addition,
  we exclude sources flagged as extended in the GB6 survey and include
  only unique pair of sources, i.e. sources in one catalog with
  multiple matches in the other catalog are excluded. Since multiple
  sidelobes of a single source may show up as different NVSS
  components, in this way we may bias the pairs toward compact
  sources. However NVSS resolution is such that less than $1\%$ of
  sources is resolved into multiple components
  \citep{2005MNRAS.362....9B}, compared with $\sim 7\%$ of GB6 sources
  with multiple NVSS matches in the reference catalog. We then
  conclude that the majority of multiple matches we find are spurious
  pairs.

 The distribution of spectral indexes so obtained describes the
 $1.4-4.85$ GHz behaviour of sources selected at the GB6 frequency,
 and can be used to accurately propagate sources from $4.85$ GHz to
 the lower frequency.  In order to use this distribution to
 extrapolate from $1.4$ GHz to higher frequencies, some adjustments are
 required. The NVSS survey has a flux limit of $2.5$mJy, while the
 GB6 includes sources above $18 $mJy, therefore low--flux NVSS
 sources which appear also in the GB6 survey will be biased to have a
 rising spectral index. Figure \ref{fig:alpha_of_S} shows the average
 $\alpha_{1.4}^{4.8}$ as a function of $S_{1.4}$. While for $S_{1.4}
 \gtrsim 0.3$ Jy the spectral index is a decreasing function of the
 NVSS flux, for $S_{1.4} \lesssim 0.1$ Jy $\alpha_{1.4}^{4.8}$ shows a
 steep upturn due to completeness issues. We therefore exclude from
 the reference sample all sources with $S_{1.4} \le 0.1$ Jy.  We
 further divide the reference sample in 10 logarithmic bins covering
 the flux range $0.1 - 3.0$ Jy, depending on the value of $S_{1.4}$,
 with an additional bin including all sources with $S_{1.4} > 3$ Jy.
 In the end the reference sample includes ~31000 sources, with corresponding
 spectral index $\alpha_i$ and uncertainty $\delta\alpha_i$.

\begin{figure}
\includegraphics[width=8.cm]{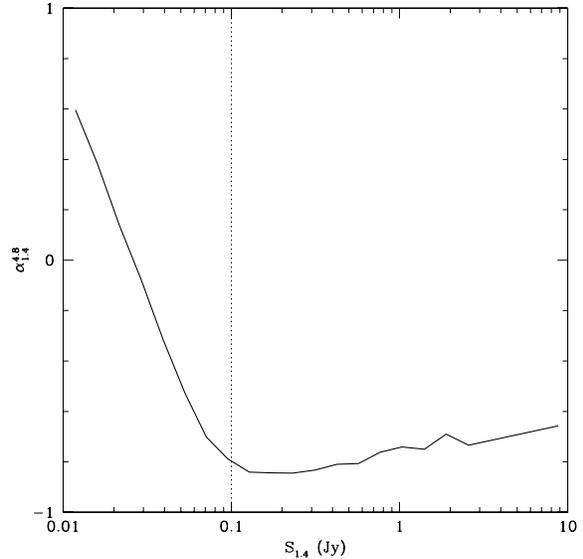}
\caption{The average $1.4 - 4.85$  GHz spectral index of GB6 sources as
 a function of $1.4$ GHz flux. Below $S_{1.4} \sim 100$mJy NVSS
 sources with a GB6 counterpart are biased to have upturn spectra.}
\label{fig:alpha_of_S}
\end{figure}

In principle, since the reference sample is based on sources with
matches in both NVSS and GB6 catalogs, we may be missing: a) sources
with strongly rising spectra, which are absent in NVSS but are
detected by GB6, or b) sources with very steep spectra, which fall
below the GB6 detection threshold. \cite{2009arXiv0908.0066W} find
that $4.3\%$ of sources selected at $15.2$ GHz with $10 {\rm mJy} \le
S_{15} \le 15$mJy do not have an NVSS counterpart. GB6 sources are
selected with higher threshold (18mJy) and at lower frequencies than
\cite{2009arXiv0908.0066W}. Therefore, sources of type (a) would
require spectral indexes significantly more rising than those of the
rising population discussed in that work, an we expect the fraction of
sources detected at $4.85$  GHz but not in the NVSS to be significantly
lower than \cite{2009arXiv0908.0066W} estimates. Regarding point (b),
we note that sources in the reference sample with $100 \le S_{1.4} \le
200$ mJy have an average spectral index $\alpha_{1.4}^{4.8} = -0.84
\pm 0.34$. Therefore we estimate the fraction of NVSS sources with no
GB6 counterpart to be $\sim 5\%$ at $100$mJy and $\sim 0.5\%$ at $150$
mJy.

\item {\it From 1.4 {\rm GHz} to 23 {\rm GHz} for sources with $S_{1.4} \ge 300$ {\rm
      mJy}.} From the previous discussion, we conclude that the
  thinned reference sample defined above provide an accurate
  description of the frequency scaling from 1.4 to 4.85 GHz for the
  population of sources with $S_{1.4} \gtrsim 300$ mJy, which are
  responsible for the bulk of source contamination at CMB
  frequencies. Therefore, each NVSS source with $S_{1.4} > 300$ mJy is
  randomly paired to a spectral index $\alpha_i$ from the
  corresponding flux bin of the thinned reference sample. We then
  randomly draw a spectral index $\bar{\alpha}$ from a Gaussian with
  mean $\alpha_i$ and width $\delta\alpha_i$. In this way we
  incorporate in the extrapolation the measured uncertainty on
  $\alpha_i$, however, since the distribution of the reference
  spectral indexes is not necessarily Gaussian, the overall
  distribution of the $\bar{\alpha}$ both within a single realization
  and among different realizations will not be Gaussian. The flux of
  the NVSS source is extrapolated to $4.85$ GHz assuming $S \propto
  \nu^{\bar{\alpha}}$. The source is then propagated to $23 $ GHz
  using the spectral indexes by \cite{2007MNRAS.379.1442W}. Since the
  flux threshold of the $15.2$ GHz sample is higher than the GB6
  limit, we need to thin the catalog to avoid that sources with low
  fluxes at $4.85$ will bias the sample toward flatter spectral
  indexes. Due to the low number of sources, we cannot study the
  behaviour of $\alpha(S)$ as done in the previous section. Instead we
  compute the average spectral index $\alpha_{4.8}^{15}$ using all the
  121 sources in the sample. Using this average $\alpha_{4.8}^{15}$,
  we extrapolate the $15.2$ GHz threshold to $4.85$ GHz and remove
  form the $4.8-15.2$ GHz reference sample all sources with $S_{4.8}$
  below this value.  The process is iterated until no more sources are
  rejected. We are left with $76$ sources with $S_{4.85} > 75$ mJy,
  which we assume provide a representative description of the
  behaviour of sources between $4.85$ and $23$ GHz. Each source
  simulated at the previous step is paired with a random source in
  this new reference sample, and its corresponding set of spectral
  index. We use this set to propagate the source to $23$ GHz, going
  through an intermediate step at $15.2$ GHz.

\item {\it From 1.4 {\rm GHz} to 23 {\rm GHz} for sources with
    $S_{1.4} \le 100$ {\rm mJy}.} For low flux sources, $S_{1.4} <
  100$ mJy, we extrapolate directly to $23$ GHz assuming $S \propto
  \nu^\alpha$, with $\alpha$ randomly drawn from the distribution of
  spectral indexes by \cite{2009arXiv0901.4330M}. In order to be
  consistent with the treatment of sources with $S_{1.4} > 100$ mJy
  described above, we assumed that such distribution be valid in the
  range $1.4 - 23$ GHz, even if it is based on $31$ GHz followup of
  NVSS sources.

\item {\it From 1.4 {\rm GHz} to 23 {\rm GHz} for sources with $100
    {\rm mJy} < S_{1.4} < 300$ {\rm mJy}.} Sources with $100 {\rm mJy}
  < S_{1.4} < 300$ mJy are above the upper validity limit of the
  \cite{2009arXiv0901.4330M} study, and may yet be slightly affected
  by spurious effects in the NVSS - GB6 reference sample defined
  above. Therefore, in this intermediate regime, we randomly chose
  between the two approaches discussed above, and tune the selection
  function to reproduce low flux counts at higher frequencies. The
  probability of extrapolating sources using the spectral index
  distribution of \cite{2009arXiv0901.4330M} is given by:
 \begin{equation}
   \label{eq:select}
   P(S_{1.4}) = \frac{1}{2} \left[1 - \tanh \left( \frac{S_{1.4} -S_*}{\Delta} \right) \right]~.
 \end{equation}
 The values $S_*$ and $\Delta$ are chosen by fitting the $33$ GHz
 number counts from the extrapolation procedure described here to the
 observed number counts by DASI and VSA in the $0.05 - 1$ Jy range.
 Sources with lower fluxes are not a significant contaminant in WMAP
 maps, as discussed above. Note that we do not require the selection
 function in equation \ref{eq:select} to be continuous at the $100$mJy
 and $300$mJy boundaries.

\item {\it From 23 {\rm GHz} to 94 {\rm GHz} for sources with $S_{23}
    \ge 1 $ {\rm Jy}.} Sources with $S_{23} > 1$ Jy are extrapolated
  to higher frequencies according to the WMAP5 catalog itself. We
  divide the WMAP5 catalog in two sub catalogs comprising sources with
  $S_{23} > 2$ Jy and $1 {\rm Jy} < S_{23} < 2$ Jy, respectively
  containing $69$ and $166$ elements. When a simulated source has
  $S_{23} > 1$ Jy, we randomly pair it with a spectral index from the
  subcatalog covering the corresponding flux range, and extrapolate
  the source to higher frequency using a simple power law.

\item {\it From 23 {\rm GHz} to 94 {\rm GHz} for sources with $S_{23}
    < 1 $ {\rm Jy}.} For fluxes below $\sim 1$ Jy the WMAP catalog
  is not complete.  \cite{2007MNRAS.379.1442W} suggest a procedure to
  extrapolate fluxes into the WMAP frequency range based on the
  spectral behaviour of the source around $\sim 20$ GHz. For sources
  for which $\alpha_{22}^{43} \le \alpha_{15}^{22} \le 0$ they suggest
  fitting a quadratic form to the $\log(S) - \log(\nu)$ relation,
  while for the remaining sources they adopt a linear $\log(S) -
  \log(\nu)$, using $\alpha_{22}^{43}$. Alternatively, we consider the
  set of spectral indexes based on the \cite{2008MNRAS.385.1656S}
  sample. Taking into account both sets of observations, we consider
  then the following extrapolation strategy for sources with $S_{23}
  <1$ Jy:

 \begin{itemize}
 \item Sources with $S_{23} < 100$ mJy are randomly paired to source
   of \cite{2007MNRAS.379.1442W} and extrapolated according to the
   procedure described there;
 \item Sources with $S_{23} > 400$mJy are propagated assuming a
   single power law with spectra index drawn randomly from the reference
   sample of \cite{2008MNRAS.385.1656S};
 \item Sources with $100 < S_{23} < 400$mJy are extrapolated choosing randomly
   between the previous two methods; the probability of selecting one method
   over the other is given by the relative number of sources in the relevant 
   flux range.
 \end{itemize}

\end{itemize}

In addition, when at a given step we have different extrapolation
procedures depending of the flux of the source, we do not switch
abruptly from one regime to the other, but we adopt a linear
transition between the two with a width which is the greater of
$100$mJy and $10\%$ of the threshold flux. Moreover, we do not
consider polarization.

\begin{figure*}
\centerline{
\includegraphics[width=14cm]{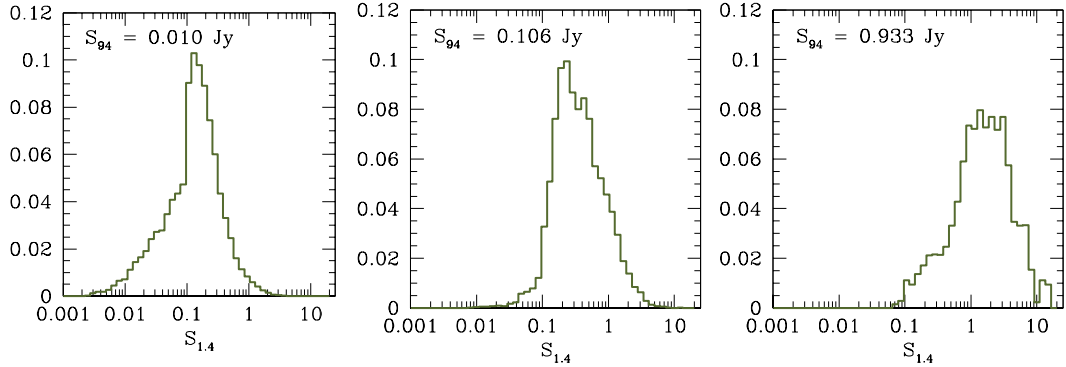}
}
\caption{Distributions of the initial 1.4 GHz fluxes for sources with a
 final 94 GHz as indicated in the respective panels. Histograms
 refer to the full set of 800 simulations and have been
 normalized to have unit area.}
\label{fig:backhist}
\end{figure*}

\begin{figure*}
\centerline{
\includegraphics[width=14cm]{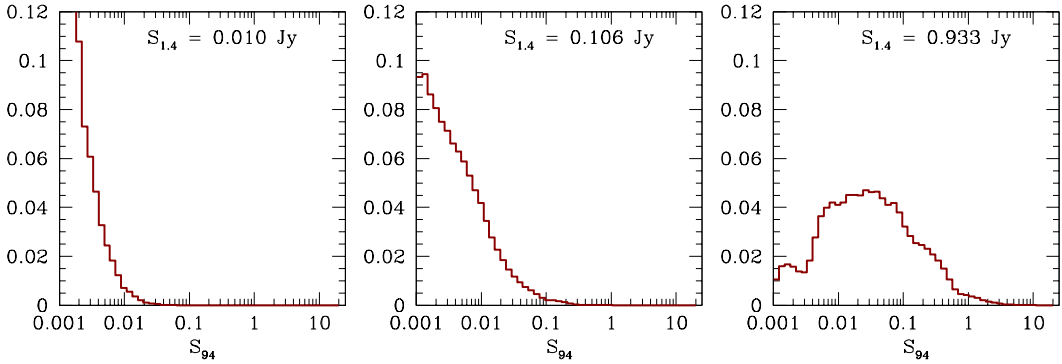}
}
\caption{As in figure \ref{fig:backhist} but showing distributions of simulated 94 GHz fluxes for sources with initial
1.4 GHz flux as indicated in the different panels.}  
\label{fig:frwdhist}
\end{figure*}

\section{Results} \label{sec:res}

\subsection{Flux distribution} \label{sec:distrib}
This approach provides both the probability that a source with an NVSS
flux $S_{1.4}$ will have a 94 GHz flux $S_{94}$,
i.e. $P(S_{94}|S_{1.4})$, as well as the complementary distribution
$P(S_{1.4}|S_{94})$.  In figure \ref{fig:backhist} we plot the
distribution of the initial 1.4 GHz fluxes for sources with final 94 GHz
fluxes of $S_{94} \sim 0.01, \sim 0.10$ and $\sim 1.0$ Jy, which
provides an estimate of $P(S_{1.4}|S_{94})$. The plots refer to the
full set of 800 simulations. Figure \ref{fig:frwdhist} shows instead
$P(S_{94} | S_{1.4})$. A general expectation is that for flux levels
relevant to CMB experiments, the high frequency population of sources
is dominated by objects with flat spectral indexes. We recover this
behaviour in the simulations, as shown in figures \ref{fig:backhist}
and \ref{fig:frwdhist}. In particular, according to the discussion of
section \ref{sec:ped}, the residual contamination in WMAP
maps will be dominated by sources with $S \gtrsim 0.1$ Jy in the
frequency range $60-90$ GHz, and we expect the majority of these
sources to have $S_{1.4} \gtrsim 100$mJy. Inaccuracy in the
extrapolations of sources with lower $S_{1.4}$ will have a minimal
impact on estimates of the residual contamination in WMAP maps.

\subsection{Number counts} \label{sec:counts} 
\begin{figure*}
\centerline{
\includegraphics[width=7.cm]{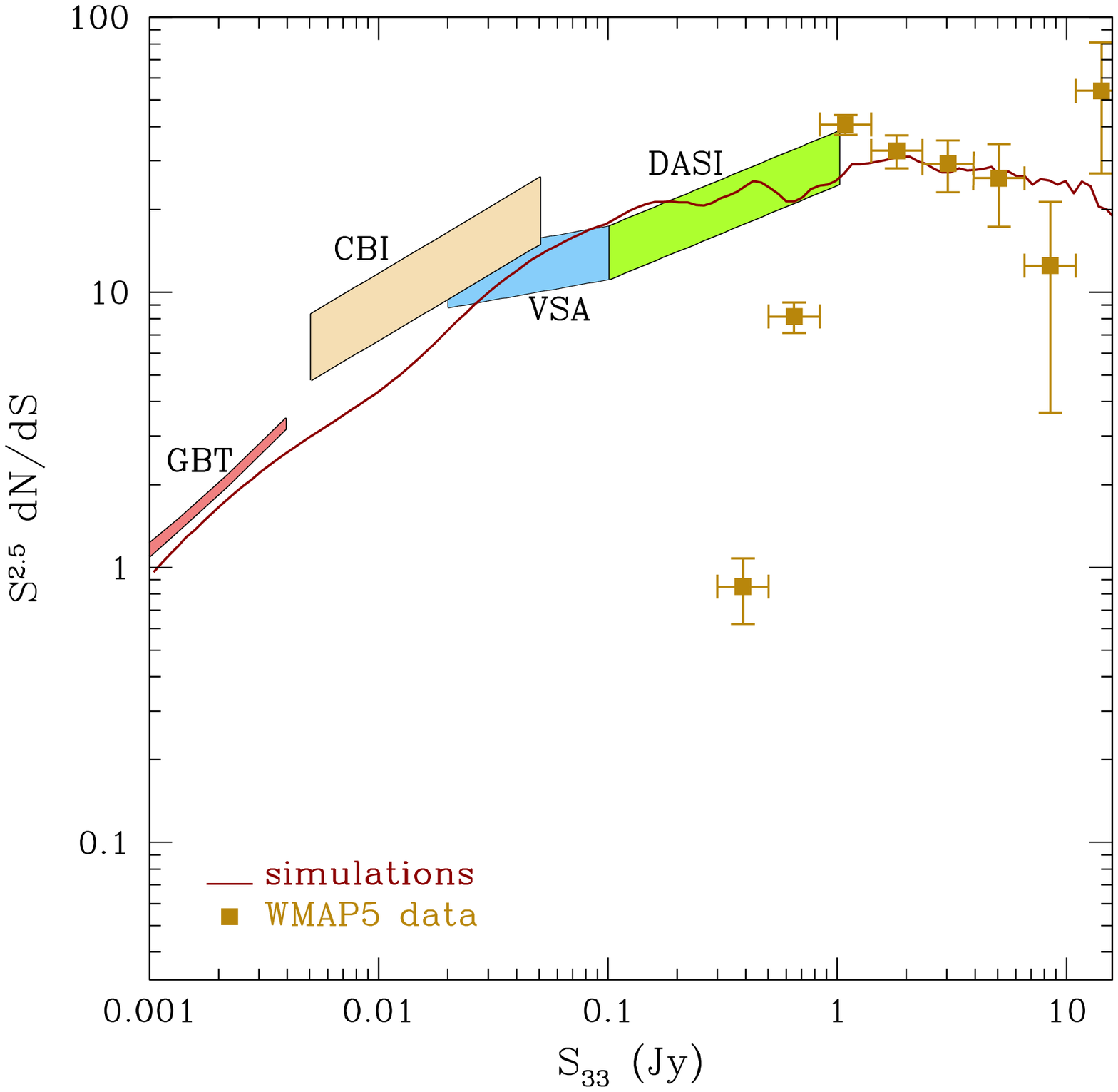}
\includegraphics[width=7.cm]{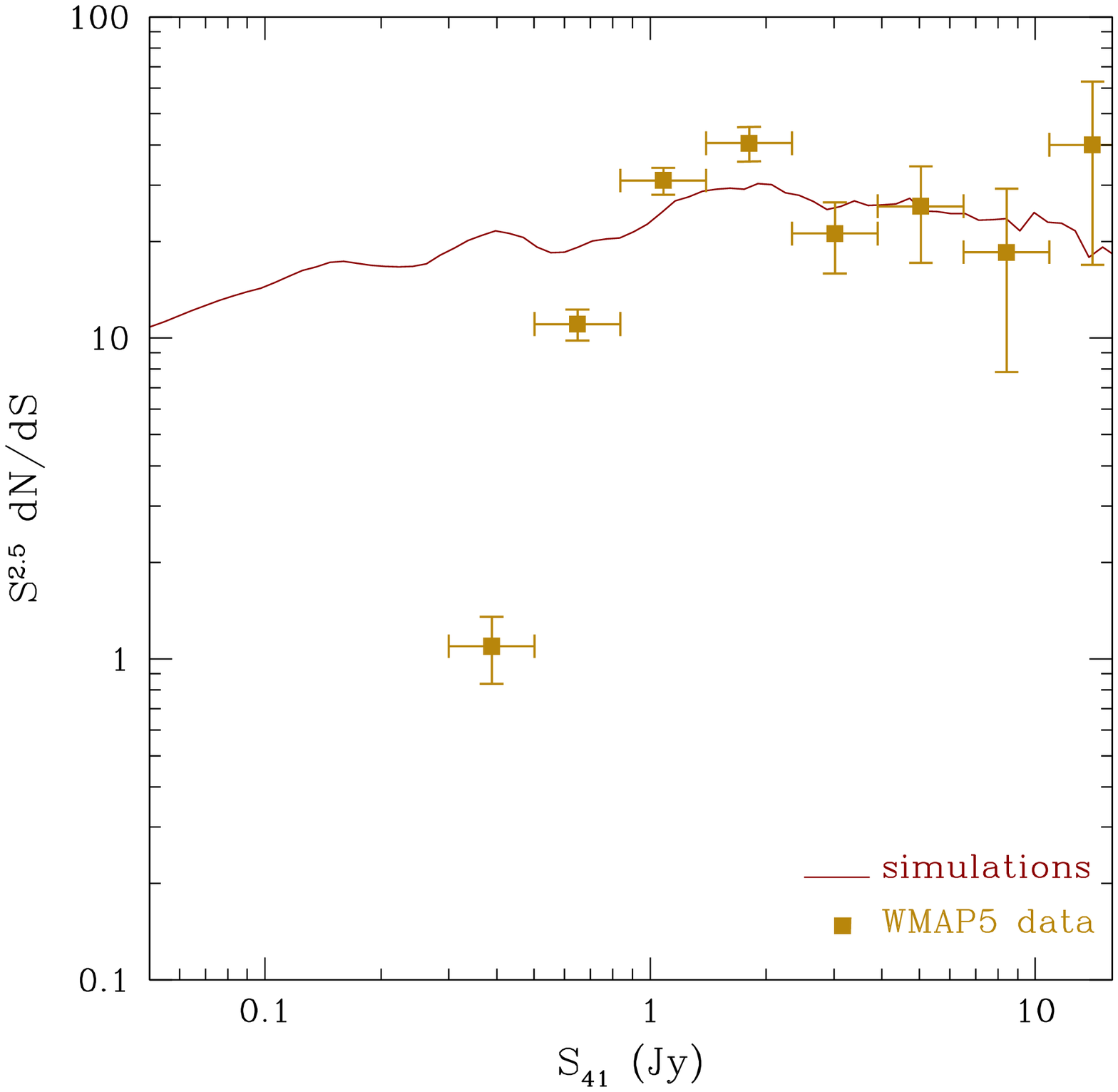}
}
\centerline{
\includegraphics[width=7.cm]{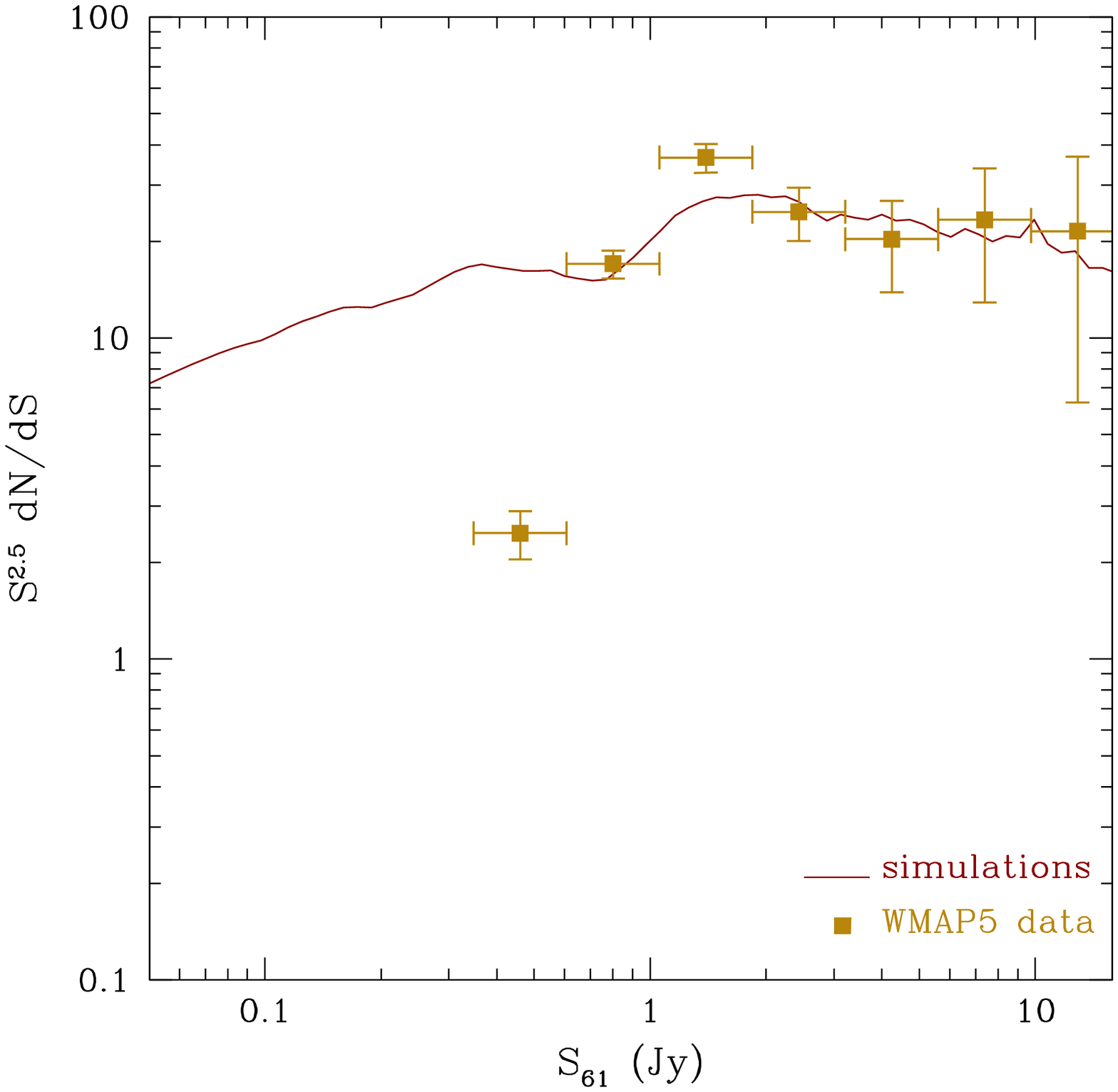}
\includegraphics[width=7.cm]{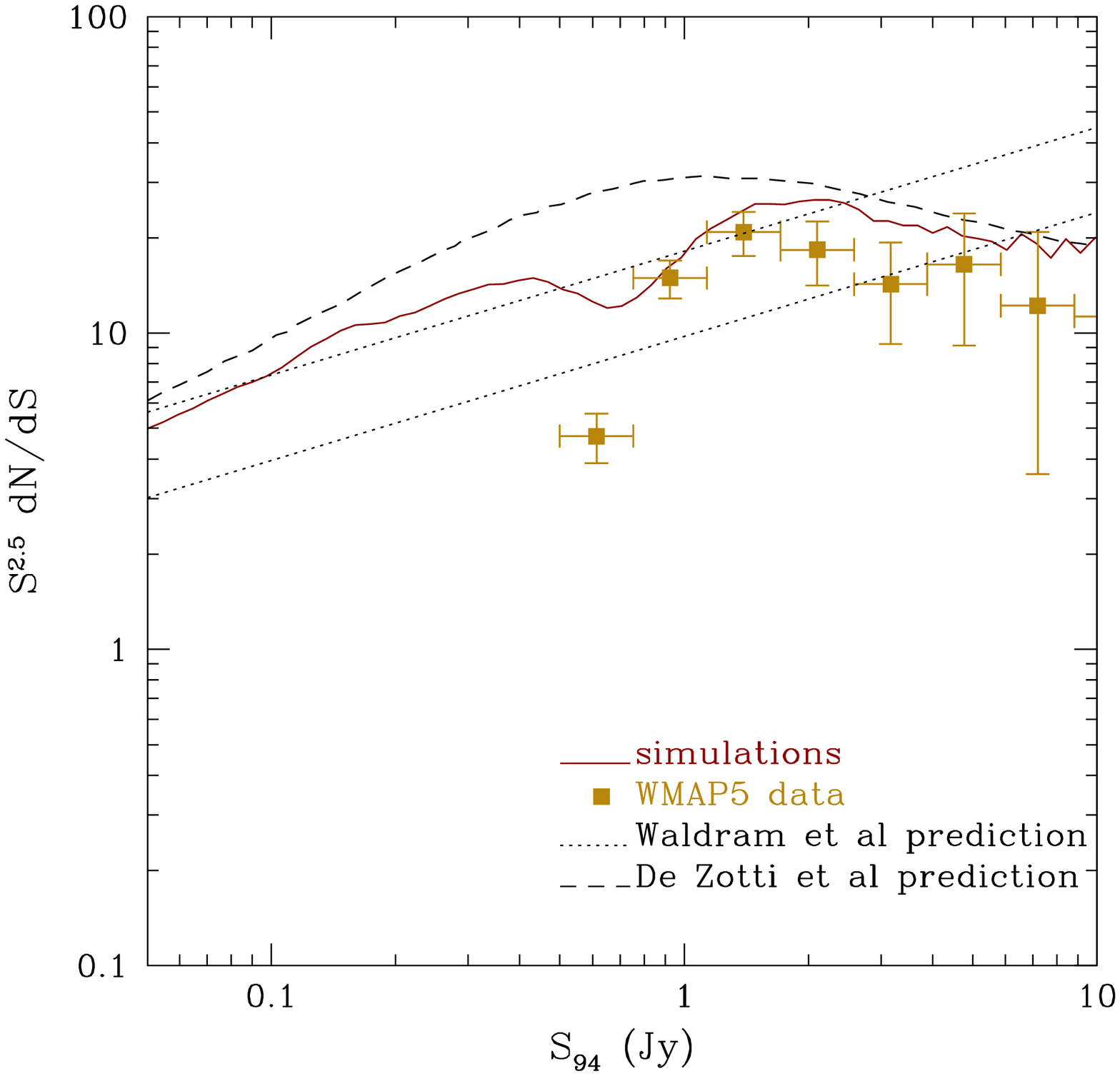}
}
\caption{Comparison of the predicted Euclidean counts from the
  ensemble of NVSS simulations of this work (solid lines) with the
  incomplete counts from the WMAP 5 year catalog (squares), at the
  frequencies of 33, 41, 61 and 94 GHz. At 33 GHz we also plot data
  from the CBI, VSA and DASI survey, while at 94 GHz we compare with
  predictions from earlier works.}
\label{fig:eucl_counts}
\end{figure*}

For each simulation, we
compute the corresponding differential number counts $dn(S)/dS$ and
then average over the whole set of simulations.  In figure
\ref{fig:eucl_counts} we compare the incomplete counts from WMAP5
source catalog with the ensemble average differential counts at the
frequencies of $33, 41, 61$ and $94$ GHz, approximately corresponding
to the central frequencies of WMAP Ka, Q, V and W bands. At 33 GHz we
compare our estimation also with results from CBI, VSA, and DASI
surveys, while at 94 GHz we also plot prediction from earlier works
~\citep{2005A&A...431..893D,2007MNRAS.379.1442W}.

Comparison with lower flux data at $33$ GHz, shows that the methods
discussed here tends to underestimate source counts below $S \sim
0.02-0.03$ Jy. These sources do not provide a relevant contribution to
the total residual contamination to WMAP spectra, but will play a more
relevant role for Planck data analysis.  Since the uncertainty in our
methods is dominated mostly by the low number of reference sources
above $4.85$ GHz, we expect a better agreement as the pool of reference
sources increases. In the range $0.5 -1.0$ Jy, simulations show
evident artifacts due to the transition between different regimes of
flux extrapolation, i.e. using spectral index from the WMAP5 catalog
for $S_{23} > 1$ Jy. They are due to the low number of sources in that
flux range in the base catalogs; as above, we expect these artifact to
significantly decrease as more data over the relevant frequency and
flux range will be included in the reference catalogs.

\begin{figure}
\includegraphics[width=8.5cm]{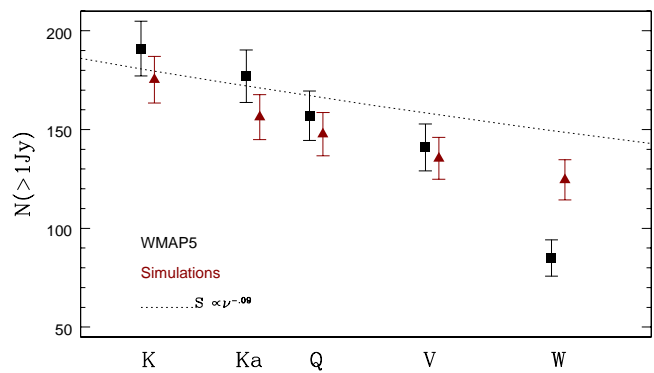}
\caption{Comparison of integrated counts in simulations
 (red/triangles) and in WMAP 5 year catalog (black/squares). The
 dashed line shows the expected behaviour assuming a scaling of
 differential counts $dn/dS \propto S^{-2.5}$ and a frequency
 dependence $S \propto \nu^{-0.09}$.  The curve amplitude is fixed by
 a fit to WMAP count in K, Ka and Q bands. In computing the number
 counts we excluded the sky area with declination $\le -40\deg$ and
 within the WMAP 5yr point source catalog mask.}
\label{fig:int_counts}
\end{figure}

While our extrapolations correctly recover WMAP5 results at lower
frequencies, counts at 94 GHz are overestimated by $\sim 50\%$. In
figure \ref{fig:int_counts} we plot the integrated counts $N(>S=1{\rm
  Jy})$ from our simulations and from WMAP5 data. For reference, we
also show an analytic estimate of the counts assuming $dN/dS \propto
S^{-2.5}$ and $S\propto \nu^{-0.09}$ \citep{2009ApJS..180..283W},
normalized to fit WMAP5 counts in K, Ka and Q bands. WMAP5 data show a
steepening of the counts in W band compared to lower frequencies,
while our counts seem to be in better agreement with a linear $\log(S)
- \log(\nu)$ relation. As discussed in section \ref{sec:methods}, our
extrapolations of sources with $S_{23} > 1$ Jy is mainly based on the
family of spectral indexes provided by the WMAP5 team, which can be
approximated by a Gaussian distribution with mean $\alpha = -0.09$ and
dispersion $\sigma_\alpha = 0.17$. Therefore, as expected source
counts in the simulations are in agreement with the analytic estimate
plotted in figure \ref{fig:int_counts}; the slightly steeper behaviour
seen in simulations is compatible with the actual distribution of
WMAP5 spectral indexes having a tail toward steeper $\alpha$. This
suggests that there is some tension between the WMAP5 W-band counts
and their spectral index distribution. A possible explanation could be
a progressive steepening of the spectral index of bright sources with
increasing frequency. \cite{2008MNRAS.384..711G} noticed that the
spectral index of WMAP sources in the $41-64$ GHz range is steeper
than their spectral index in the $5-23$GHz range, and radio source
measurements in the $\sim 100 -250$ GHz range by the QUaD telescope
\citep{2009ApJ...700L.187F} and the South Pole Telescope
\citep[SPT,][]{2009arXiv0912.2338V} provide additional evidence in
this direction \citep[see][for a recent review]{2009A&ARv.tmp...15D}.
If this is actually the case, describing the source behaviour in the
$23 -94$ GHz range with a single power law may not be entirely accurate.

Alternatively, there may be some unaccounted for systematics affecting
WMAP detection efficiency in W band. The WMAP team required that for a
source to be included in the catalog it needed to be detected at more
than $5\sigma$, in at least one band. Its flux in the other bands
would be included if: a) it was measured at more than $2\sigma$, b)
the fitted source width is within a factor of 2 of the actual beam
\citep{2009ApJS..180..283W}. \cite{2009arXiv0912.0524S} pointed out
that the W band beam response to discrete sources appears to depend in
a non-linear fashion on source flux. In particular, they find that the
effective beam profile is significantly wider for point sources than
for the Jupiter observations used as a basis for the WMAP team's beam
analysis, which can create a significant bias in determining if a
source fits requirement (b). Since number counts from simulations do
not include systematics due to WMAP actual detection procedure, this
effect, if confirmed, could be an alternative explanation for the
discrepancies between simulated and actual number counts. Further data
on source behaviour at high frequency, e.g. from Planck, will help
solve this issue.

In addition, all WMAP sources with measured W band flux have been
detected also in at least one of the lower frequencies.  As long as a
WMAP source is detected in at least one band, it is masked in the
power spectrum analysis \citep{2009ApJS..180..296N}, and we follow the
same procedure in building sky mask for the analysis of section
\ref{sec:spectra}. Therefore, any mismatch in the number of W band
bright sources between simulations and real data will not impact our
estimates of the unresolved source power as long as counts at the
lower frequency for the simulations are in good agreement with WMAP
findings.

\section{Maps} \label{sec:maps}

An advantage of this approach consists in that it naturally provides
an estimates of bright sources position and accounts for clustering of
unresolved sources. However, agreement of number counts estimated from
simulations with those computed by data does not guarantee that the
method correctly recovers the spatial distribution of sources.  In
order to check whether the simulations match the structure observed in
WMAP maps, for each realization we produce source-only (i.e. without
noise and CMB) maps and compare them with WMAP observed maps. Source
maps also allow and independent determination of unresolved source
contamination. Since WMAP estimate of this contribution is based on
the Q, V and W bands data, we generate map for each of the above
frequencies.

We place the sources on an
HEALPIX\footnote{{http://healpix.jpl.nasa.gov/}}\citep{gorski:05} map at
a resolution ${\rm N}_{\rm side} = 2048$, corresponding to a pixel
size of $\sim 1.7'$.  This value is considerably greater than the NVSS
beam and angular resolution, so we can consider the sources as points
when placing them on the map, and considerably smaller than WMAP W
beam of $\sim 13'$, so that any artifacts due to pixelization,
smoothing etc. will be minimal. The actual WMAP beams are not
Gaussian, their exact profile varying between each set of Differential
Arrays (DAs). To reduce the number of maps we have to deal with, we
consider individual frequencies, rather than individual DAs. Each map
is then smoothed with an effective beam given by the average, in
harmonic space, of the beam profiles of the DAs of the corresponding
frequency. The smoothed maps are then degraded to a resolution ${\rm
N}_{\rm side} = 512$.

In order to check that our predictions recover the point source structure
seen in the maps, we compute the linear correlation coefficient
between each individual simulation and the corresponding
WMAP 5yr map:
\begin{equation}
\label{eq:r}
r = \frac{\langle \left( {\bf X}_i -\langle {\bf X} \rangle \right) \,
\left( {\bf M}_i -\langle {\bf M} \rangle \right) \rangle_{S > S_c}}
{\sigma_{\bf X} \, \sigma_{\bf M}}~.
\end{equation}
Here ${\bf X}$ and ${\bf M}$ respectively refer to the simulated and
WMAP maps, after application of an harmonic space filter
$b_\ell/(b_\ell^2 C_\ell + C_\ell^{\rm noise})$, where $b_\ell$ is the
harmonic transform of the beam profile, $C_\ell$ is a theory CMB power
spectrum based on WMAP5 best fit parameters
\citep{2009ApJS..180..306D} and $C_\ell^{\rm noise}$ is the noise
power spectrum \citep{2009ApJS..180..283W}. The means and variances of
the maps, $\langle{\bf X}\rangle$ and $\sigma_{\bf X}$, and the
correlation coefficients are computed using only pixels within radius
equal to the nominal frequency FWHM, of sources with flux above a
threshold $S_c$. To assess the significance of this result, we compare
the value of $r$ computed from the simulations-WMAP5 correlations,
with that computed by correlating our simulations with a mock
microwave sky which includes contribution from CMB, WMAP5-like noise,
Galactic foregrounds and a point sources population obtained according
to the procedure described in this work.  In practice, we randomly
chose one of the simulations as the actual distribution of point
sources, and add to this a CMB realization, an anisotropic Gaussian
noise term and a foreground contribution. The pixel noise variance is
given by $n = n_0 /\sqrt{N_{\rm obs}}$, where $N_{\rm obs}$ is the
effective number of observations in each pixel and $n_0 = 2.197$ mK
for the Q band and $6.547$ mK in W band \citep{limon}. However, we do
not account for noise correlations between different pixels. The
foreground contribution is based on the Maximum Entropy Method maps
\footnote{{http://lambda.gsfc.nasa.gov/product/map/dr3/mem\_maps\_info.cfm}}
\citep{2009ApJS..180..265G}. The maps are provided at a resolution
$N_{\rm side} =128$ and smoothed with a $1\deg$ beam, and are thus have no
power at scales $\ell \gtrsim 300$. Therefore the foregrounds
contribution to the variance and mean of the mock sky, in particular
to the small scales relevant to point sources, will be significantly
lower than the impact on the actual WMAP sky maps. We generate
different sets of mock skies, corresponding to different point
sources, CMB and noise realizations. The mock sky then constitute a
best case scenario, in which the point source distribution follows
exactly the model discussed in this work, and several systematic
effects, like foregrounds and correlated noise, are absent or play a
reduced role. Tables \ref{tab:corr_Q} and \ref{tab:corr_W} show
results of the comparison in Q and W bands, respectively.

\begin{table*}
\centerline{
\begin{tabular}{@{}lccccccc}
\hline 
$S_c$ (Jy) & WMAP5 & mock  & mock  & mock  & mock & mock  & mock \\
          & maps  & data 1& data 2 & data 3 & data 4 & data 5 & data 6\\
\hline
1.0 & $0.22 \pm 0.07$ & $0.34 \pm 0.08$ & $0.36 \pm 0.07$ & 
$0.37 \pm 0.08$ & $0.35 \pm 0.07$ & $0.32 \pm 0.07$ & $0.37 \pm 0.07$ \\
0.5 & $0.18 \pm 0.05$ & $0.30 \pm 0.06$ & $0.32 \pm 0.05$ & 
$0.31 \pm 0.06$ & $0.29 \pm 0.05$ & $0.26 \pm 0.05$ & $0.31 \pm 0.05$\\
0.2 & $0.13 \pm 0.04$ & $0.23 \pm 0.04$ & $0.25 \pm 0.04$ & 
$0.23 \pm 0.04$ & $0.22 \pm 0.04$ & $0.19 \pm 0.03$ & $0.25 \pm 0.04$\\
0.0 & $0.03 \pm 0.02$ & $0.06 \pm 0.02$ & $0.07 \pm 0.01$ &
$0.06 \pm 0.01$ & $0.06 \pm 0.01$ & $0.05 \pm 0.01$ & $0.08 \pm 0.01$\\
\hline
\end{tabular}
}
\caption{Linear correlation coefficient $r$ between simulated source maps at 
 $41$  GHz and WMAP 5 year Q band maps, for different flux cuts $S_c$. 
 Only pixels within 1FWHM radius of sources with $S>S_c$ are included in the 
 computation of $r$. For comparisons, columns 3-8 show the same quantity 
 computed by correlating our set of simulations with mock WMAP-like skies 
 for which the point source component follows the same distribution as our 
 simulations.}
\label{tab:corr_Q}
\end{table*}

\begin{table*}
\centerline{
\begin{tabular}{@{}lccccccc}
\hline 
$S_c$ (Jy) & WMAP5 & mock  & mock  & mock  & mock & mock  & mock \\
          & maps  & data 1& data 2 & data 3 & data 4 & data 5 & data 6\\
\hline
1.0 & $0.14 \pm 0.08$ & $0.21 \pm 0.09$ & $0.24 \pm 0.08$ & 
$0.20 \pm 0.08$ & $0.18 \pm 0.07$ & $0.16 \pm 0.07$ & $0.18 \pm 0.07$ \\
0.5 & $0.14 \pm 0.06$ & $0.18 \pm 0.07$ & $0.21 \pm 0.06$ & 
$0.16 \pm 0.06$ & $0.15 \pm 0.05$ & $0.14 \pm 0.05$ & $0.15 \pm 0.05$\\
0.2 & $0.12 \pm 0.04$ & $0.13 \pm 0.04$ & $0.16 \pm 0.04$ & 
$0.11 \pm 0.03$ & $0.11 \pm 0.03$ & $0.11 \pm 0.03$ & $0.11 \pm 0.03$\\
0.0 & $0.014 \pm 0.003$ & $0.011 \pm 0.003$ & $0.014 \pm 0.003$ &
$0.010 \pm 0.002$ & $0.010 \pm 0.002$ & $0.010 \pm 0.003$ & $0.011 \pm 0.002$\\
\hline
\end{tabular}
}
\caption{As table \ref{tab:corr_Q} but for W band.}
\label{tab:corr_W}
\end{table*}

{A general remark is that even in the best case scenario we expect
 only a moderate level of correlation: $r \lesssim 0.40$ and $r
 \lesssim 0.25$ for bright sources, respectively in Q and W band. As
 expected, the degree of correlation between simulations and actual
 maps is lower than that between simulations and mock data. While for
 W band the difference is within the associated standard deviation,
 in Q band the discrepancy is of order $2 \sigma$ for bright sources
 increasing to $\sim 3 \sigma$ for faint ones. Foreground
 contamination is more relevant at $41$  GHz than at $94$  GHz and this
 may account for the significant loss of correlation in Q band.

\section{Power spectra and effect on parameters}\label{sec:spectra}

The WMAP team estimated the amplitude of the unresolved point source
contribution to a cross spectrum $C_\ell^{Y_1Y_2}$ ($Y_1 Y_2$
represents a pair of different DAs, e.g. $Q_1 Q_2$, $Q_1 V_1$) assuming
a power-law scaling in frequency:
\begin{equation}
\label{eq:source_cl}
C_\ell^{\rm src} = A_0 g(\nu_1) g(\nu_2) \left( \frac{\nu_1}{\nu_Q} \right)^\alpha
\left( \frac{\nu_2}{\nu_Q} \right)^\alpha,
\end{equation} 
where $\nu_1, \nu_2$ are the frequencies of the DAs considered,
$g(\nu)$ was defined in equation \ref{eq:gnu}, and
the spectral index $\alpha$ is assumed to be either 0, as most sources
are expected to have a flat spectrum, or -0.09, corresponding to the
mean of of a Gaussian fit to the spectral indexes of the actual WMAP5
catalog \citep{2009ApJS..180..283W}. The common amplitude $A_0 = 0.011
\pm 0.001 \, \mu{\rm K}^2 {\rm sr}$ is determined by fitting the
spectral shape of equation \ref{eq:source_cl}, to the cross spectra
from the Q,V,W bands; the estimated $A_0$ has only a small dependence
on the two values of $\alpha$ considered
\citep{2009ApJS..180..296N}. The final power spectrum, instead, is a
linear combination of the possible VV,VW, and WW cross spectra,
weighted by the corresponding inverse covariance matrix
\citep{2003ApJS..148..135H,2009ApJS..180..296N}. This method is
internal, as it includes information only from WMAP own measurements.
On the contrary, the approach discussed in this paper, provides a
partially external method, in the sense that it incorporates both
information from WMAP data (to propagate sources with $S_{23} > 1$ Jy)
and from other surveys, and could provide a cross check of WMAP
results.

\begin{figure*}
\includegraphics[width=14.cm]{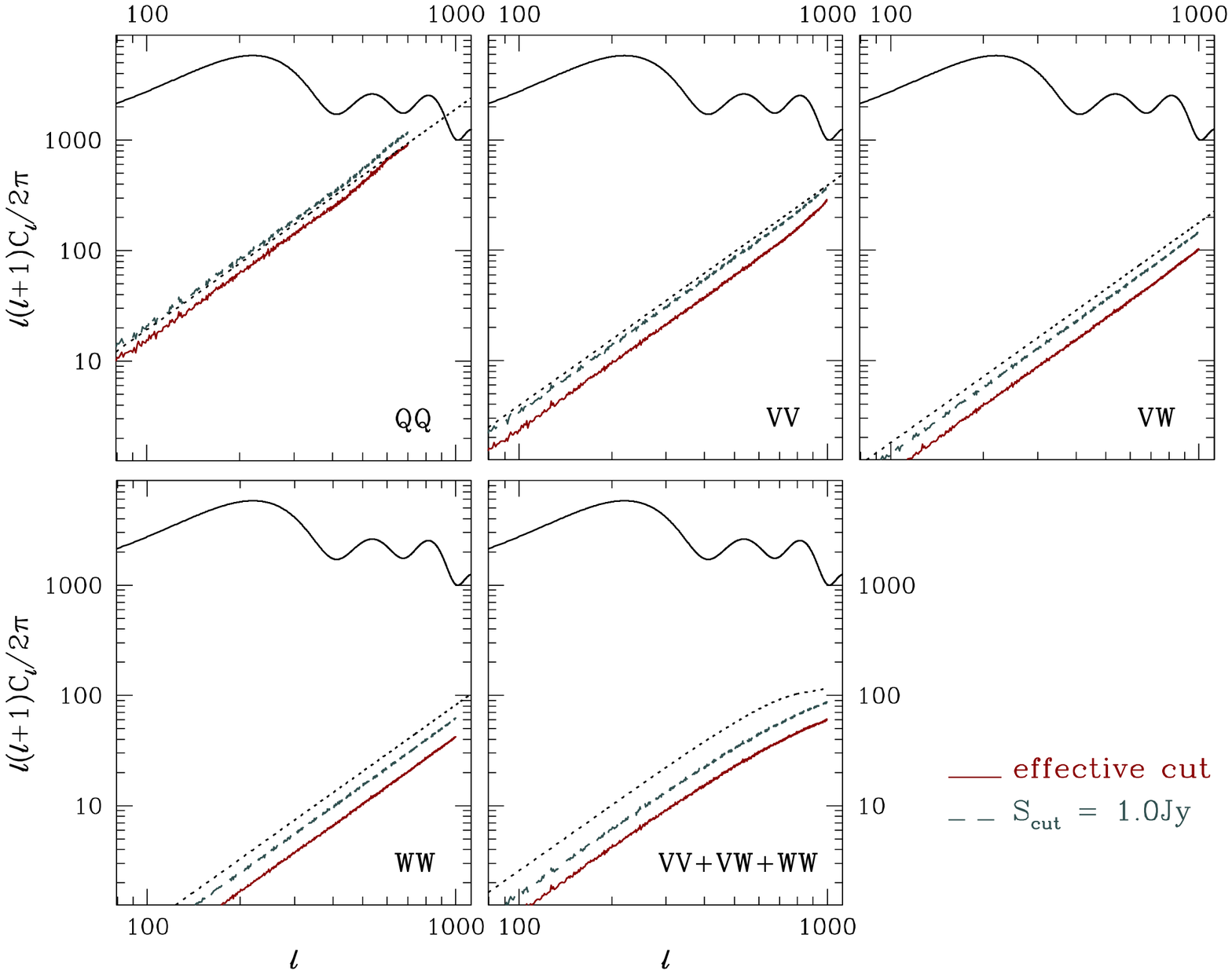}
\caption{Estimates of the unresolved point sources contribution to CMB
 power spectra at different WMAP frequencies when applying the
 masking procedure described in the text (solid lines) or simply
 removing all sources with Q band flux $S_{41} > 1 Jy$ (dashed lines).
 The dotted lines show a Poisson spectrum with
 normalization $C_\ell^{src} \propto A_0 \nu_1^\alpha \nu_2^\alpha$
 with $A_0 = 0.011$ and $\alpha = -0.09$. The final power WMAP power
 spectrum is a combination of VV, VW and WW cross spectra and does
 not depend on QQ data. For reference we also show the CMB
 temperature power spectrum for the best fit WMAP5 cosmological
 model.}
\label{fig:spectra}
\end{figure*}

To estimate the unresolved point sources contamination from our
simulations we need to define an individual mask for each
realization. We start by applying the same Galactic cut as for the
KQ85\footnote{{http://lambda.gsfc.nasa.gov/product/map/dr3/masks\_info.cfm}}
mask, used in the WMAP5 power spectrum analysis
\citep{2009ApJS..180..296N}, and also mask the area outside of the
NVSS sky coverage. This base mask is the same for all realizations.
In addition, the KQ85 mask removes the sky area within $0.6 \deg$ of
sources in the WMAP5 catalog. The WMAP detection efficiency for radio
sources depends on a number of factors, including frequency, position
on the sky and the actual CMB and noise fluctuations around the
source.  Replicating all these effects in all our simulations would be
too time consuming and instead we adopt a simplified procedure to
``detect'' and mask sources. The WMAP5 catalog is complete for sources
$\gtrsim 1$ Jy at all frequencies, and the number counts for those
sources follow an Euclidean distribution.  Therefore, at each
frequency we assume a power law shape for the number counts $F_{\nu}
S^{-2.5}$, and fix the value of $F_{\nu}$ by fitting to the observed
WMAP5 differential number counts ${\bar n}_\nu(S)$ for $S > 1$ Jy at
the appropriate frequency. We then assign to each simulated source a
detection probability $P(S_*) = {\bar
 n}_\nu(S_*)/(F_{\nu}S_*^{-2.5})$. On average, in each simulation we
detect $304 \pm 16$ sources, compared with 321 actual WMAP5 sources in
the same sky area, and we mask a $0.6 \deg$ radius around each
detected source.  Alternatively, we consider the conservative approach
of masking only sources with a Q band flux $S_{41} > 1.0$ Jy.

We apply the resulting mask to the corresponding set of maps and
compute the QQ, VV, VW and WW spectra using a MASTER approach
\citep{2002ApJ...567....2H}. The VV, VW and WW spectra are then
combined according to \cite{2003ApJS..148..135H}; notice that we only
include the diagonal terms of the covariance matrix. The final
estimate of the unresolved sources contribution is the average of the
spectra from the individual realizations.

In figure \ref{fig:spectra} we compare our estimates with WMAP
results. While our predictions are in good agreement with WMAP5 yr
estimates for the QQ spectra, they show a steeper frequency
behaviour. In particular, for the final combined power spectrum, we
predict a residual point source contribution $\sim 50 \%$ lower than
WMAP estimates for the source detection strategy discussed above, and
$\sim 25\%$ lower if we only mask sources with $S_{41} > 1 Jy$.
Uncertainty on simulations results is $\sim 10\%$, comparable to the
error on $A_0$. Notice that in figure \ref{fig:spectra} we do not plot
our estimates of residual sources contamination at $\ell > 700$ for
the Q band, as the uncertainty in the beam asymmetries for that channel
introduce significant artifacts at $\ell \gtrsim 500$
\citep{2009ApJS..180..246H}. Since we do not use the Q band data to
estimate the point sources contribution to the final spectra, these
artifacts do not alter the following discussion.

The differences between our prediction and WMAP5 estimates arise from
a combination of factors. As discussed above, the reference catalog we
use to extrapolate high flux sources from $20$ GHz to $94$ GHz was
given by the subsample of WMAP5 sources with with fluxes $S_{23} > 1$
Jy, corresponding to the approximate completeness limit of WMAP 5yr
catalog. WMAP sources with $S_{23} > 1$ Jy have an average spectral
index ${\bar \alpha} = -0.21 \pm 0.30$, while for sources with $S_{23}
< 1$ Jy ${\bar \alpha} = -0.09 \pm 0.66$. The spectral index used by
the WMAP team in order to derive the amplitude is the mean of a
Gaussian fit to the spectral index distribution of all their detected
sources, and is therefore affected by the shallower spectral indexes
of sources with $S_{23} < 1$ Jy. Note that we did not use the latter
set of sources for the simulations, as for this range of fluxes and
frequencies we adopted the results of \cite{2008MNRAS.385.1656S}.
These results suggest that sources with fluxes in the $0.2 - 1$ Jy
range have an average spectral index ${\bar \alpha} = -0.43 \pm 0.31$,
while for the subset of 8 sources with $S_{20} > 0.8$ Jy, ${\bar
  \alpha} = -0.26 \pm 0.16$, in good agreement with estimates for WMAP
sources with $S_{23} > 1$ Jy. Our simulations reflect this finding. As
more information on the spectral behaviour of sources with K band flux
in the $0.5 - 1.5$ Jy will become available, it will be possible to
better establish if the low flux sources of WMAP catalog are biasing
the estimates of the spectral index used for estimating the unresolved
point sources contribution.

In addition, WMAP fix the amplitude of the unresolved source
contribution $A_0$ by a simultaneous fit to the ensemble of cross
spectra obtained using Q, V and W bands, assuming a constant frequency
scaling across the corresponding range of frequencies, 41 - 94
GHz. Therefore, although the final CMB power spectra are a combination
of only V and W measurements, the estimated unresolved sources
contribution to such spectra depends also on Q data. In this work,
instead, we estimate the point source contamination at a given
frequency directly from the point sources maps at that frequency,
without including information from other bands. In particular, our
prediction for the final combined VV+VW+WW spectrum does not depend on
41 GHz data.}  As discussed in \cite{2009ApJS..180..296N}, the WMAP5
estimated value of $A_0$ using only V and W data is $\sim 10\% -40\%$
(depending on the choice of Galactic mask) lower than when using Q,V
and W data, although the uncertainty in the former case is $\sim 3$
times the uncertainty in the latter. Therefore, using information from
Q band may lead to a slight overestimate of $A_0$ at $60 -90$ GHz.

As pointed out by \cite{2006ApJ...651L..81H,2008ApJ...688....1H} the
point sources correction and the way it is treated in the likelihood
function affects the determination of the cosmological parameters, in
particular of the slope of the power spectrum of scalar perturbations,
$n_s$.  To test the effect, if any, of our approach on parameter
estimation, we use COSMOMC\footnote{http://cosmologist.info/cosmomc/} \citep{2002PhRvD..66j3511L}  + PICO\footnote{http://cosmos.astro.uiuc.edu/pico/} \citep{2007ApJ...654....2F} to run a set of Monte-Carlo Markov
Chains (MCMC) replacing the source correction term in WMAP5 likelihood
code with our results. However, we did not change the shape of the
likelihood function as suggested by \cite{2008ApJ...688....1H}.  We
consider the six base $\Lambda$CDM parameters: physical baryon,
$\omega_b$, and cold dark matter, $\omega_c$, densities; optical depth
to reionization, $\tau_e$; tilt, $n_s$, and amplitude, $A_s$ of the
power spectrum of scalar density fluctuations; the angle subtended by
the sound horizon at recombination, $\theta$. We include
marginalization over the amplitude of the Sunayev-Zeldovich effect and
lensing. As expected, since changing the source correction effectively
alters the shape of the second and third acoustic peaks in the
measured spectrum, we find that $\tau$ and, to a lesser degree,
$\theta$ and $A_s$ are unaffected. For our default source removal
procedure, the shift on $n_s$ and $\omega_c$ can reach almost
$0.4\sigma$, with a noticeable shift also for the baryon density,
$\omega_b$, while if we adopt a more conservative masking, the effect
can be up to $0.3 \sigma$.

\begin{figure}
\includegraphics[width=9.cm]{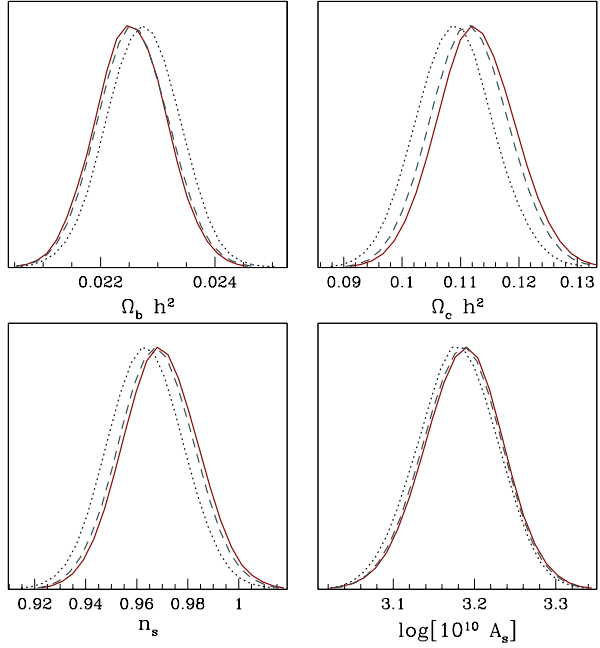}
\caption{Dependence of parameter estimates on the shape of the source
 correction. Curves refer to default WMAP 5yr estimates (black/dotted
 lines), or our prediction when masking sources as described in the
 text (red/solid) or simply masking sources with $S_{41} > 1$ Jy
 (green/dashed).}
\label{fig:params}
\end{figure}

\section{Summary and conclusions}\label{sec:conc}

We discussed a new method aimed at evaluating the residual point source 
contribution to the WMAP5 power spectrum. We implemented a
stochastic approach to extrapolate the flux of radio
sources of the NRAO-VLA Sky Survey (NVSS) to WMAP frequencies. Using
different point source catalogs with multifrequency information, we
divided the $1.4 - 90$ GHz in a number of frequency steps. At each step
an NVSS source is randomly paired to a reference source in the catalog
covering the relevant frequency range, and extrapolated to the upper
limit of that range using the spectral index of the reference source.
The procedure is iterated 800 times, and the ensemble of
simulations provides the probability distribution for the flux each
NVSS at the desired frequencies.

We compared the statistical properties of the simulations with those
of the sources of the WMAP5 catalog, with other surveys at 33 GHz and
with predictions by previous authors at 94 GHz, finding an agreement
down for fluxes $\gtrsim 0.1$ Jy in Ka band, while we underestimate
the number of sources below $\sim 0.1$ Jy. The WMAP catalog is not
complete below $S \sim 1$ Jy, so point sources with $S \sim 0.1$ Jy
account for only a small fraction of the total power due to unresolved
sources. At higher frequencies predictions are in general agreement with 
previous works.

An advantage of the method is that it maintains the information about
the sources position. In order to test this, we considered the sets of
point-source-only (i.e. no CMB nor noise) maps obtained from NVSS
simulations at 41 and 94 GHz, and correlated them with both the Q and
W band WMAP5 sky maps and with mock CMB skies in which the point
source population follows the same distribution as our model. We found
that although the overall level of correlation is low, W band
correlations with actual data are in good agreement with correlations
with the mock data. In Q band, instead, the degree of correlation
between simulations and actual maps is significantly lower than what
expected from the mock data. In this band foregrounds contamination is
significantly higher than at higher frequency and may contribute to
decreasing the level of correlation between simulations and real
maps.

We used the simulated maps to estimate the unresolved source
correction to WMAP5 measured spectra. As a general result, we find
that unresolved sources in our simulations have a steeper frequency
behaviour than the contribution estimated by the WMAP team. This is
due to the fact that observation of sources in the 20--90  GHz range
with $S \lesssim 1$ Jy typically show steeper spectral indexes than
the best fit spectral index from the WMAP5 catalog. The observed
scaling in the simulations corresponds to an effective spectral index
$\alpha = -0.20 \sim -0.25$, compared to WMAP5 $\alpha = 0 \sim
-0.09$. Thus, the unresolved point sources contribution derived by the
simulation either agrees with the WMAP5 estimates in Q band but is
lower by up to $\sim 30-40\%$ at 94 GH, depending on the way we mask
bright sources. The corresponding shift in estimates for parameter
like $n_S$ and $\omega_c$ can reach up to $\sim 0.3-0.4\sigma$.

Given a sufficient number of simulations, the approach followed in
this work allows in principle to estimate the probability distribution
for the flux at an arbitrary frequency of each source from a template
low frequency survey. In turn this would allow to naturally account
for the added uncertainty due to undetected sources during the
map-making stage, as is done for galactic foregrounds with Gibbs
sampling methods. This would require that Gibbs samplers can be made
to efficiently work at multipoles $\ell \simeq 2000-3000$, where point
sources will be relevant for Planck and other possibly upcoming
high-resolution experiments. Even if this were not possible, having an
estimate of unresolved source contributions in each pixel would be
helpful in defining a sky mask, e.g. by flagging all pixel in which
the source contribution is expected to exceed a given threshold even
if there is no actual detection of a point source in that region of
the map.

\section*{Acknowledgments}

EP is and NSF--ADVANCE fellow (AST--0649899).
 LPLC and EP were supported by NASA grant
NNX07AH59G and Planck subcontract 1290790 for this work, and would
like to thank Caltech for hospitality during this period. We
acknowledge the use of the LAMBDA data archive and of the CosmoMC,
PICO and HEALPix packages. We thank the members of the US Planck ADG
team for helpful and stimulating discussion.

\clearpage


\end{document}